**Fusion energy commercialization requires solving social and environmental challenges**

Stephanie Diem[1], Laila El-Guebaly[2], and Aditi Verma[3]
[1]sjdiem@wisc.edu | [2]laila.elguebaly@wisc.edu | [3]aditive@umich.edu

**Abstract**
Fusion energy, the process that uses the same reaction that powers the sun and the stars, offers the promise of virtually unlimited, carbon-free energy and is approaching reality. Recently, there's been a dramatic global increase in the investment and research focused on addressing the hurdles to commercialize fusion energy. While a majority of the effort has been focused on gaps in technology, little work has been done to address the societal and environmental impacts of this technology. Three community- and environmentally-focused research priorities are identified for commercializing fusion energy: 1) understanding the environmental impacts of fusion energy across the technology lifecycle, 2) developing risk and safety assessment methodologies for fusion power plant technologies, and 3) creating a community-based socially engaged approach for fusion technology design and development. This approach will benefit private companies who wish to deploy future fusion power plants as concerns about the technology will be addressed early in the design process, thus minimizing delays in deployment that may result in increased costs for developers. Community engagement around fusion technology development must be evidence- based in order to build trust between communities and technology developers. Such an approach is grounded in informed consent is vital for the sustainable development and use of fusion technologies.

**Main Article**
Combating the looming threat of an already changing climate requires swift global action to develop clean energy solutions and decarbonize energy systems with the limited resources available on our planet.

Fusion energy, the process that uses the same reaction that powers the sun and the stars, offers the promise of virtually unlimited, carbon-free energy and is approaching reality. While the main application of fusion energy is for electricity generation, the fusion reaction can also supply process heat for carbon-free industrial applications. Recent major advancements in creating the conditions for fusion have resulted in the Joint European Torus (JET) experiment more than tripling the previous record of energy produced during sustained magnetic confinement fusion energy experiments [Gibney 2022] and the breakthrough achievement of ignition at the National Ignition Facility (NIF) in 2022 [Nature 2022]. Technological advances, like high-temperature superconducting magnets that are now capable of providing a path to more compact fusion power plants, have motivated a historic pivot from a field traditionally focused on fundamental science research to one that addresses technical challenges needed to sustainably harness energy from fusion, as noted in a recent National Academies study [NASEM 2021]. The emergence of fusion technology-focused startups and companies, and record-breaking private investments in fusion globally (a total of $6 billion with more than $4 billion since 2020) [FIA 2023] add further momentum to the commercialization of fusion energy, potentially in time to avert the worst effects of climate change and energy infrastructure renewal, while supporting economic development. In 2022, the U.S. White House organized a Summit on Developing a Bold Decadal Vision for Commercial Fusion Energy followed later that year by the launch of a Milestone-Based Fusion Development Program by the U.S. Department of Energy to provide federal support for commercial development of fusion energy to design fusion pilot plants capable of generating electricity. More recently in November 2023, an international engagement plan to advance fusion commercialization was announced at COP28 to focus on research and development, building fusion supply chains, as well as regulation and safety standards so that fusion energy technologies may be made available worldwide. These initiatives signal to governments around the world to mobilize tools and infrastructure to pursue an aggressive and ambitious timeline for fusion energy development, placing commercial fusion energy within reach in a manner that was previously not deemed possible by researchers, practitioners, and policymakers. In addition, particularly in the U.S., federal officials are placing an emphasis on ensuring that the benefits of fusion are shared in a just and equitable way during the process. These are areas that have previously received insufficient attention in the research and practitioner communities in fusion and broader energy sectors.

---

[1] Assistant Professor of Nuclear Engineering and Engineering Physics, University of Wisconsin-Madison.
[2] Distinguished Research Professor - Emerita, Engineering Physics Department, University of Wisconsin-Madison.
[3] Assistant Professor of Nuclear Engineering and Radiological Sciences, University of Michigan.

Fusion energy has historically been positioned in contrast to all other sources of energy as a clean, environmentally friendly, free of long-lived radioactive waste, and potentially unlimited source of energy. Many of these claims about a perfect, environmentally benign source of energy lack critical analysis and must be supported by rigorous research. Creating the conditions for and sustaining the fusion reaction requires materials that can withstand the most harsh environments in the universe (10-20 times hotter than the core of the sun), both in temperature and radiation. Research on the environmental impact of fusion power plants suggests that the sizable radioactive waste (radwaste) volumes should be addressed especially as these wastes though not as long-lived, will be many times larger than waste volumes generated by fission reactors. Recycling and clearance, which allows for removing materials that are not contaminated or activated from a fusion power plant site and no longer controlling them by a regulatory body, considerations need to be incorporated early into the design process [El-Guebaly 2021].

Additionally, researchers and developers will need to consider the social and environmental impacts of mining resources, such as lithium and beryllium, as well as considerations of sourcing conflict minerals such as tantalum and tungsten, used to build and fuel a fleet of future fusion power plants. Finally, land and water use and impacts of fusion energy are also not well understood.

Many fusion technology developers have been pursuing their technology commercialization efforts on the assumption that (as yet unproven) minimal societal and environmental impact confers on them a mandate to develop technology without community consultation and engagement. The history of all energy technologies is marked by the creation and entrenchment of environmental injustices resulting from exclusionary technology development and deployment practices. For example, across nearly all energy technologies – nuclear fission, solar, wind – the socio-environmental impacts of resource extraction – mining of uranium, copper, zinc, nickel, are disproportionately borne by indigenous and socioeconomically disadvantaged communities who are often displaced by the siting of mines or have to bear the health burdens and environmental costs of mining activities. Such communities have also typically hosted waste disposal facilities.

Energy technology siting practices have historically not accounted for the preferences of the communities that host these technologies with community consultation being limited or too late in the technology development process for communities to be able to offer meaningful input.  In other words – the benefits and burdens resulting from the development and use of energy technologies have not been equitably distributed. These inequitable technology development practices and the resulting public opposition are likely to become a significant impediment to the large-scale decarbonization of our energy systems [Xu 2020].

Fusion energy technologies and their designers, being in the early stages of development while the system designs are the most flexible, can learn from and avert the mistakes made in the development of other clean energy technologies.  Despite the challenges that lie ahead, fusion energy has the potential to provide energy security to nations as the most widely pursued approach relies on an abundantly available fuel –  heavy forms of hydrogen as fuel. Additionally, fusion can provide baseload electric power as well as process heat to decarbonize industrial processes. It is the most dense form of energy, providing 4 times more energy per kilogram of fuel than nuclear power and 4 million times more energy per kilogram of coal or oil.

No energy technology is perfectly equitable and the development of any new technology – energy or otherwise – results in benefits and burdens. Building on previous calls to work towards social acceptance of fusion technology [Wurzel and Hsu 2022; Hoedl 2022], we propose a sociotechnical development approach for fusion energy that centers community engagement from the earliest stages of technology development.

We propose three guiding principles for commercializing fusion energy:
(1) <u>Centering communities:</u> Acknowledge that fusion technologies are the means to an end – climate and energy security – and that the impacted communities and their concerns must be centered in efforts to commercialize the technology. This can be done through direct engagement in the early stages of technology development.
(2) <u>Identifying and understanding risks:</u> Pursue a rigorous program of research to characterize and understand the gaps in our knowledge about the safety, security, and proliferation risks and environmental impacts of commercial fusion technology. Early community engagement may help identify these gaps, as well as prioritize their importance to communities who may host fusion facilities.

(3) <u>Anticipating and addressing social and environmental concerns:</u> Having understood the risks and environmental impacts of commercial fusion technology, as well as the concerns from communities, we must analyze and anticipate these socio-economic-environmental benefits and burdens and their distribution, in order that the fusion community can preemptively and proactively learn about and respond to equity and justice issues.

Community engagement around fusion technology development must be evidence-based in order to build trust between communities and technology developers. Such an approach grounded in informed consent is vital for the sustainable development and use of fusion technologies. Since fusion technology developers are envisioning global markets and applications of their technologies, these engagement efforts must ideally be undertaken on an international scale.

**Three main areas of research**
We propose three areas of research in support of a sociotechnical approach to fusion technology development. These are:

1. **Understanding the environmental impacts of fusion energy across the technology lifecycle**
There's a need to understand the environmental impacts of fusion technology development across the fusion energy system lifecycle. This includes mining requirements, environmental impacts of fusion power plants where they are sited, as well as an assessment of the types of radioactive and non-radioactive waste generated by fusion power plants and options for their interim storage, land-based disposal, recycling, and clearance [El-Guebaly 2018]. Life cycle assessment (LCA) for fusion energy has been limited to assessing $CO_2$ emissions across the lifecycle of tokamaks [Tokimatsu 2000]. The focus on tokamaks is due to their more mature design point compared to other fusion energy system designs as well as the availability, in some cases, of full system designs.  A critical LCA expanding the entire lifecycle beyond $CO_2$ impact is necessary for fusion due to the unique conditions necessary for fusion reactions to occur on Earth: extremely high temperatures (up to 10-20 hotter than the sun) and a higher energy (14.1 MeV) neutron environment, more extreme than what is encountered in nuclear fission reactors, leading to larger volumes of low-level radwaste than produced by fission. Research to develop reduced-activation materials and radiation-resistant structural and functional materials that can withstand this environment is a critical challenge for fusion energy. Design choices for these materials with strict alloying elements and impurity control will determine the environmental impacts from resource mining, fabrication process, amount of radwaste (and its classification), and the ability of these materials to be recycled or cleared. Additionally, the impacts of tritium breeding material choices and handling of tritiated materials during recycling and radwaste management will need to be understood. It is important to understand this environmental life cycle impacts across different fusion power plant technologies as well as for design variations within a particular fusion concept  (tokamak, stellarators, spherical tokamak, etc.). Traditional LCA focuses on the environmental impacts of energy technologies, and assessments should expand to social life cycle assessment to focus on impacts on communities as well [Grubert

2018]. Future efforts incorporating a dynamic life cycle assessment and assessments of inequities in technology deployment can consider a variety of aspects such as displaced and new jobs and technologies during a rapid and large-scale energy transition [Edward 2014]. Having understood the broader impacts from a social life cycle and equity assessment for a variety of fusion energy system designs, it is important to consult with communities who might one day host such fusion facilities, develop a shared understanding and expectation of environmental impacts and how to manage them, and informed by this knowledge, develop fusion-specific regulations.

**2. Developing risk and safety assessment methodologies for fusion power plant technologies**

Though fusion power plants will not experience core meltdowns that have been a serious cause of concern for large light water fission reactors in particular and are therefore, by comparison, inherently safer, it is nevertheless important to understand the safety concerns associated with building and operating them. Previous work on fusion safety was primarily carried out in the U.S. since the early 1980s. More recently, fusion safety work applying probabilistic risk assessments [DOE Standard 1996] has been carried out by a few pro-fusion countries due to the lack of good data for component failure rates and reliability for many novel systems. Given the national interest in rapidly developing and commercializing fusion power plants, it is important to: (1) update previous safety studies for new technologies and design concepts, (2) evaluate the suitability of a range of possible safety assessment methodologies – probabilistic risk assessments being one potential methodology, (3) learn about community concerns about fusion safety as well as establish a socially acceptable level of safety, and (4) develop detailed guidelines for the regulatory application of these safety and risk assessment methodologies, considering that fusion has different radionuclide profile with fewer risk issues compared to fission. The use of large amounts of tritium and high-energy neutrons in fusion energy systems pose unique radiation risks to the environment and human health. Security and proliferation risks of different fusion energy technologies must also be assessed to determine whether these risks can be best addressed through technology design or institutional and normative controls. An opportunity exists now to develop new risk and safety assessment approaches that do not rely on a full system design, thereby providing an opportunity to assess the safety, security, and proliferation risks without fully designed systems. This approach can be applied to a general approach to fusion, i.e. magnetic confinement fusion while considering common components and points of failure. Incorporating community concerns early on during this process provides another path to build trust and transparency early in the development phase. Pursuing this novel approach will allow for a more robust path forward to embed environmental and energy equity early into system design while the designs are the most flexible.

**3. Creating a socially engaged approach for fusion technology design and development**
Both research areas above highlight the importance of engaging with communities from the earliest stages of technology development and assessment. Such participatory and community-engaged approaches have been applied for the development of consumer products and systems and for the assessment of science-based initiatives such as solar geoengineering

[Kaplan et al. 2019 ] and NASA's asteroid initiative [Tomblin et al. 2017], but not for the design and assessment of larger, more complex sociotechnical systems [Verma & Allen 2022 ].³ Applying adaptations of these methods to fusion energy development can provide insight on community members' current understanding of fusion during two-way meaningful engagements while preparing them to engage, either individually or to identify trusted representatives or organizations to represent them, in participatory design or co-design. Additionally, by coupling these engagements with opportunities to join initiatives such as the U.S. DOE Clean Energy Communities or longitudinal survey studies for several months post-engagement with communities, empowers community participants to stay involved during the design process. While a unique approach like this can be focused on a specific technology such as fusion energy, lessons learned in the application can be adapted to other emerging technologies as they are developed in the early phase of development. A broader impact of this approach is that we can train the next generation of technology developers to engage in two-way engagement with community members [Martin 2023]. This is a vital third area of research for fusion technology development. Absent a socially engaged approach to technology development, private fusion energy companies and the federal government risk making massive investments (already numbering in the billions) in and pursuing the development of technologies that may not be accepted by communities. A socially engaged technology development approach is not only the right thing to do but is also the strategically and economically superior option. Researchers working in the advanced fission area are developing community-engaged and participatory approaches to technology development. Operationalizing such an approach in a fusion context requires the creation of novel design and development processes, tools, and technology assessment frameworks – all of which must be supported by rigorous research.

Fusion has the potential to be adapted globally as the energy source to power the next phase of human technology and civilization. By pursuing the approach described here and centering the three guiding principles for fusion energy above, fusion energy technologies and their designers can learn from and avert the mistakes made in the development of other energy technologies. To support pursuing this necessary research that centers environmental justice and energy equity to develop fusion energy for all of humanity, it will require a transdisciplinary approach to engage a wide variety of expertise such as community engagement, risk communication, environmental studies, risk and safety, and policy experts.

**References**
1. 2022 Gibney, "Nuclear-fusion reactor smashes energy record", Nature 602, 371, https://doi.org/10.1038/d41586-022-00391-1
2. 2022 Lawrence Livermore National Laboratory, "National Ignition Facility achieves fusion ignition", Press release - https://www.llnl.gov/news/national-ignition-facility-achieves-fusion-ignition
3. 2022 Tollefson and Gibney, "Nuclear-fusion lab achieves 'ignition': what does it mean?", Nature 612, 597-598 (2022), https://doi.org/10.1038/d41586-022-04440-7

---

³ The transactions paper is behind a paywall. Here is an easily accessible copy
https://drive.google.com/file/d/17molBaWNqf4WveYF2FS4dpLy1IcENkPW/view?usp=share_link

4. 2021 NASEM, "Brining Fusion to the U.S. Grid", https://www.nationalacademies.org/our-work/key-goals-and-innovations-needed-for-a-us-fusion-pilot-plant
5. 2023 Fusion Industry Association - The Global Fusion Energy Industry in 2023 report
6. 2021 El-Guebaly - Activated Fusion Radwaste Disposition/Recycling/Clearance presented at the ARPA-E GAMOW kickoff meeting, January 2021
7. 2020 Q. Xu et al, "Structural conflict under the new green dilemma: Inequalities in development of renewable energy for emerging economies", Journal of Environmental Management, Volume 273, 111117 - https://www.sciencedirect.com/science/article/pii/S0301479720310446
8. 2022 Wurzel and Hsu, "Progress towards fusion energy breakeven and gain as measured against the Lawson criterion", Physics of Plasmas 29, 062103, https://aip.scitation.org/doi/10.1063/5.0083990
9. 2022 Hoedl, "Achieving a Social License for Fusion Energy", Physics of Plasmas 29, 092506, https://aip.scitation.org/doi/10.1063/5.0091054
10. 2018 El-Guebaly, ARIES Team and FNSF Team, "Nuclear Assessment to Support ARIES Power Plants and Next-Step Facilities: Emerging Challenges and Lessons Learned", Fusion Science and Technology 74, no. 4: 340. https://doi.org/10.1080/15361055.2018.1494946
11. 2000 K. Tokimatsu, et al, "Evaluation of $CO_2$ emissions in the life cycle of tokamak fusion power reactors", Nuclear Fusion 40, 653. https://iopscience.iop.org/article/10.1088/0029-5515/40/3Y/328/pdf
12. 2018 E. Grubert, "Rigor in social life cycle assessment: improving the scientific grounding of SLCA", The international Journal of Life Cycle Assessment, Volume 23, pages 481-491 - https://link-springer-com.ezproxy.library.wisc.edu/article/10.1007/s11367-016-1117-6
13. 2014 M.R. Edwards and J.E. Trancik, "Climate impacts of energy technologies depend on emissions timing", Nature Climate Change 4, 347-352 - https://www.nature.com/articles/nclimate2204
14. 1996 DOE Standard, Safety of Magnetic Fusion Facilities: Guidance, DOE-STD-6003-96. Currently under revision: https://www.standards.doe.gov/standards-documents/6000/6003-astd-1996/@@images/file
15. 2019 Kaplan, Leah, et al. "Cooling a warming planet? Public forums on climate intervention research." *Consortium for Science, Policy & Outcomes*. Arizona State University Washington, 2017 DC. https://cspo.org/wp-content/uploads/2019/10/SRM_book_EPUB.pdf
16. 2017 Tomblin, David, et al. "Integrating public deliberation into engineering systems: Participatory technology assessment of NASA's Asteroid Redirect Mission." *Astropolitics* 15.2 141-166. https://www.tandfonline.com/doi/pdf/10.1080/14777622.2017.1340823
17. 2022 Aditi Verma and Todd Allen, "A sociotechnical readiness level scale for the development of advanced nuclear technologies". Proceedings of the International High Level Radioactive Waste Management Conference (IHLRWM) https://www.ans.org/meetings/wm2022/sessions/attachment/paper-6409/version-2/

18. 2022 M.J. Martin, et al, "The climate is changing. Engineering education needs to change as well", Journal of Engineering Education, 28 September 2022 - https://onlinelibrary.wiley.com/doi/full/10.1002/jee.20485